\documentclass[aps,preprint,tightenlines,nofootinbib,11pt,longbibliography,superscriptaddressm,notitlepage]{revtex4-1}
\usepackage{setspace}
\usepackage{dcolumn}   
\usepackage{bm}        
\usepackage{amssymb}   
\usepackage{xcolor}
\usepackage[normalem]{ulem}
\usepackage{amsmath}
\usepackage{mathtools}
\usepackage[english]{babel}
\usepackage{accents}
\usepackage{natbib}
\usepackage{enumitem}
\usepackage[bottom]{footmisc}
\usepackage[section]{placeins}
\usepackage{dsfont}
\usepackage{natbib}
\usepackage{relsize}
\usepackage{setspace}

\usepackage{etoolbox}
\patchcmd{\section}
  {\centering}
  {\raggedright}
  {}
  {}
\patchcmd{\subsection}
  {\centering}
  {\raggedright}
  {}
  {}
  
  \usepackage{mciteplus}
    \usepackage{hyperref}

\newcommand{\de}{\mbox{d}}
\newcommand{\be}{\begin{equation}}
\newcommand{\ee}{\end{equation}}

\newcommand{\pa}{\partial}

\newcommand{\plus}{\scriptscriptstyle +}
\newcommand{\minus}{\scriptscriptstyle -}
\newcommand{\scrscr}{\scriptscriptstyle}

\newcommand{\GFTbounce}{Oriti:2016qtz,*Oriti:2016ueo}
\newcommand{\aether}{Jacobson:2000xp,*Jacobson:2010mx}
\newcommand{\deHaro}{deHaro:2018hiq,*deHaro:2018sqw}

\mciteSetSublistMode{f}

\numberwithin{equation}{section}

\begin{document}

\title{Limiting curvature mimetic gravity for group field theory condensates}

\author{Marco de Cesare}
\email{marco.de\_cesare@unb.ca}
\affiliation{Department of Mathematics and Statistics, University of New Brunswick, Fredericton, NB, Canada E3B 5A3}

  \begin{abstract}
 Nonsingular bouncing cosmologies are realized in limiting curvature mimetic gravity by means of a multi-valued potential $f(\Box\phi)$ depending on the d'Alembertian of a scalar field $\phi$. We determine the functional form of such a potential so as to exactly reproduce the cosmological background dynamics obtained in the group field theory approach to quantum gravity. The original proposal made by Chamseddine and Mukhanov~[JCAP {\bf 1703}, no. 03, 009 (2017)], which was shown to lead to a background dynamics reproducing the effective dynamics of loop quantum cosmology, is here recovered as a particular case for some specific choices of the parameters of our model. We also clarify some issues related to the multi-valuedness of the function $f$, showing in particular that its functional form can be unambiguously reconstructed by imposing appropriate matching conditions at the branching points.

\end{abstract}
  
\maketitle

\newpage


\tableofcontents


\section{Introduction}
Spacetime singularities are among the most striking predictions of classical general relativity 
with matter fields satisfying energy conditions~\cite{Hawking:1969sw}. 
Their occurrence signals the breakdown of general relativity as a valid description of the dynamics of the gravitational field in extreme regimes, such as encountered in the interior of black holes and at the earliest stages of cosmic expansion. The initial singularity problem is not alleviated in the inflationary scenario, which still yields a past-incomplete spacetime \cite{Borde:2001nh}. Thus, new physics is required in order to explain the initial conditions of our Universe.

There is a widespread expectation that spacetime singularities can be resolved in quantum gravity. Indeed, progress in background independent approaches has provided increasing evidence for such an expectation. In particular, in loop quantum cosmology (a symmetry-reduced version of loop quantum gravity) the initial singularity is replaced by a smooth bounce \cite{Ashtekar:2011ni}.  
More recently, nonsingular bouncing cosmologies have been obtained from the dynamics of condensate states in group field theory \cite{\GFTbounce}, and in quantum reduced loop gravity \cite{Alesci:2016xqa}, thus hinting at the possibility that singularity resolution is indeed achieved in full loop quantum gravity.
While these results are highly encouraging, an effective field theory incorporating non-perturbative effects for {\it all} modes of the gravitational field (beyond the homogeneous one) is not available at present.
In this regard, valuable insights can be gained from modifications of general relativity capturing the essential features of quantum gravity models \cite{Bodendorfer:2017bjt}. Such modified gravity theories can then be regarded as  toy models for a full-fledged effective theory of quantum gravity, and therefore can be used as a testing ground to derive observational consequences of Planckian effects.

A modification of general relativity resolving the initial singularity in cosmology has been recently proposed in Ref.~\cite{Chamseddine:2016uef}. The theory is an extension of the scalar-tensor theory known as mimetic gravity~\cite{Chamseddine:2013kea,Chamseddine:2014vna}, and is obtained by including higher-order derivatives in the scalar sector of the gravitational action. 
More specifically, the action includes a potential $f(\Box\phi)$ depending on the d'Alembertian of a scalar field $\phi$ satisfying the so-called mimetic constraint. Despite the presence of higher-order derivatives in the action, the theory contains only one propagating scalar degree of freedom as a consequence of the mimetic constraint. In fact, mimetic gravity belongs to a larger class of theories, known as {\it degenerate higher-order scalar-tensor} (DHOST) theories \cite{Liu:2017puc,Langlois:2018jdg} (in turn a generalization of {\it beyond Horndeski} theories), which are not plagued by the Ostrogradsky instability (see also \cite{Langlois:2018dxi} for a review) .

 Upon closer inspection, the potential $f$ turns out to be a function of the expansion of an irrotational congruence of timelike geodesics; the latter can be identified with the world-lines of mimetic dark matter and singles out a privileged (albeit dynamical) frame.
In Ref.~\cite{Chamseddine:2016uef}, a specific choice for $f$ is made that reproduces the effective dynamics of loop quantum cosmology (LQC) in the cosmological sector. The relation with LQC has been investigated in detail in Refs.~\cite{Liu:2017puc,Bodendorfer:2017bjt,Bodendorfer:2018ptp}. However, more general background evolutions can be realized by making different choices.
In particular, singularity resolution can be achieved in this context by making a suitable choice for 
$f$, such as to implement the idea of limiting curvature. 
However, it is in general not enough to assign the functional form of a single-valued $f$; 
in fact, as shown in Ref.~\cite{deHaro:2018cpl} for general bouncing cosmological models, $f$ must be multi-valued in order to consistently describe the evolution of the background. 
Care is then required in order to ensure suitable regularity at the branching points that can guarantee a smooth evolution of the cosmological background. 

In this paper we construct a mimetic gravity theory that exactly reproduces the background evolution obtained from group field theory condensates in Ref.~\cite{\GFTbounce}. Group field theory provides a second quantized reformulation of loop quantum gravity \cite{Oriti:2013}; in this approach, cosmological dynamics is emergent and can be obtained by considering a particular class of quantum states (condensates) encoding data that are associated to homogeneous geometries \cite{Gielen:2013naa,Oriti:2015qva}. 
The background dynamics obtained from group field theory condensates \cite{\GFTbounce} generalizes the effective dynamics of LQC, which can indeed be recovered as a special case (see also \cite{Adjei:2017bfm,Wilson-Ewing:2018mrp}). In this work we determine the multi-valued function $f$ corresponding to such a background evolution and analyze its properties in detail. We focus in particular on its behaviour around the branching points and derive asymptotic formulas that are valid in such regimes (namely at the bounce and at maximum expansion rate). By requiring a smooth evolution of  the cosmological background through the branching points, we are able to determine $f$ up to a divergence term. The function $f$ can be uniquely determined by further requiring that it be symmetric under the exchange of the expanding and contracting branches. 

The plan of the paper is as follows.  
Section~\ref{Sec:Mimetic} provides a brief review of mimetic gravity and its cosmology, which is included in order to make the paper self-contained; experts in the field may wish to skip to the next sections where our new results are presented. In this section, we discuss in particular the physical and geometrical interpretation of the mimetic constraint, pointing at connections with other models. The homogeneous and isotropic sector is then analyzed, and we provide a description of the ensuing cosmological dynamics by means of effective fluids; the latter represent corrections to the standard Friedmann dynamics introduced by the $f$-term.  In Section~\ref{Sec:Relational} we reformulate the dynamics of the cosmological sector of mimetic gravity in relational terms; to be specific, we introduce a minimally coupled massless scalar field playing the role of a matter clock. This is needed as a preliminary step in order to compare the background dynamics with that of group field theory, since the latter is naturally formulated in relational terms. Section~\ref{Sec:GFTbackground} briefly reviews the background evolution in group field theory and sets the stage for our applications.
In Section~\ref{Sec:GFT} we derive the function $f$ corresponding to group field theory, which represents the main result of our work. In Section \ref{Sec:Asymptotics} we study the behaviour of such function near the branching points, namely in a neighbourhood of the bounce and at the extrema of the Hubble rate. The results thus obtained are then used in Section~\ref{Sec:Matching} to determine the values of the integration constants by imposing matching conditions between the different branches. The LQC case is analyzed in Section~\ref{Sec:LQC}; we show that the corresponding multi-valued function $f$ can be obtained from our more general result by setting one of the parameters of the model to zero. In particular, the matching conditions are unambiguously and uniquely determined. We conclude with a discussion of our results in Section~\ref{Sec:Conclusion}.

\

{\bf Conventions}
An overdot indicates differentiation w.r.t.~proper time, whereas a prime denotes derivative w.r.t.~the scalar field $\psi$ (matter clock).  
Units are chosen such that $8\pi G=1$. The metric $g_{\mu\nu}$ has  signature $(+---)$. Indices are lowered and raised using the metric.

\section{Limiting Curvature Mimetic Gravity}\label{Sec:Mimetic}
Mimetic gravity was first proposed in Ref.~\cite{Chamseddine:2013kea} in order to mimic the effects of cold dark matter within the context of modifications of general relativity. In its original formulation, the theory assumes the Einstein-Hilbert action principle as a starting point and relies on the identification of the conformal factor of the physical metric with the gradient-squared of a scalar field, i.e.~upon introducing an auxiliary metric $\tilde{g}_{\mu\nu}$, one has for the physical metric $g_{\mu\nu}=(\tilde{g}^{\alpha\beta}\pa_\alpha\phi \pa_\beta\phi)\tilde{g}_{\mu\nu}$. The dynamics is then determined by extremising the action w.r.t.~the auxiliary metric and $\phi$. It was then shown in Ref.~\cite{Golovnev:2013jxa} that such a factorisation of the metric amounts to a restriction of the original variational principle; an equivalent formulation of mimetic gravity was then provided, showing that the dynamics of Ref.~\cite{Chamseddine:2013kea} can be obtained from the Einstein-Hilbert action supplemented by an extra term enforcing the mimetic constraint (\ref{EQ:MimeticConstraint}).

Further modifications of mimetic gravity were considered in Ref.~\cite{Chamseddine:2016uef}, where a new term $f(\Box\phi)$ depending on the d'Alemebertian of the scalar field  was included in the gravitational Lagrangian.\footnote{The simplest model with $f(\chi)$ quadratic was considered earlier in Ref.~\cite{Chamseddine:2014vna}.} In particular, the idea of limiting curvature can be implemented in mimetic gravity by means of a suitable choice for the newly introduced function $f$, which allows for singularity resolution in cosmology \cite{Chamseddine:2016uef} and black holes \cite{Chamseddine:2016ktu}.\footnote{An alternative realization of the limiting curvature hypothesis by means of higher order curvature invariants was studied in Ref.~\cite{Yoshida:2017swb} (see also earlier works \cite{Brandenberger:1993ef,Mukhanov:1991zn,Easson:1999xw}). The existence of a maximal curvature scale (which could be a priori independent of the fundamental one $\sim\ell_{\rm Pl}^{-2}$) was first hypothesized by Markov \cite{markov1982limiting,markov1987possible}, and the consequences of such an assumption were further analyzed in Refs.~\cite{Frolov:1988vj,Frolov:1989pf,Ginsburg:1988jq}. }\textsuperscript{,}\footnote{For a derivation of the black hole solution of Ref.~\cite{Chamseddine:2016ktu} from a Hamiltonian perspective, see Ref.~\cite{BenAchour:2017ivq}.}

\subsection{Action Principle and Interpretation of the Mimetic Constraint}

Limiting curvature mimetic gravity is a scalar-tensor theory with the action \cite{Chamseddine:2016uef}
\be\label{Action}
S[g_{\mu\nu},\phi,\lambda,\psi]=\int\de^4x \sqrt{-g}\; \left(-\frac{1}{2}R+\lambda(g^{\mu\nu}\pa_\mu\phi\pa_\nu\phi-1)+f(\chi)\right) +S_{\rm m}[g_{\mu\nu},\psi]~.
\ee
The first term is the standard Einstein-Hilbert action. Note that $\chi$ is not an independent field, being defined as $\chi=\Box\phi$. The function $f(\chi)$ is in general multivalued; in fact, this is necessary in order to accommodate bouncing backgrounds \cite{deHaro:2018cpl}.
We have included a matter action $S_{\rm m}$, with $\psi$ denoting a generic matter field. It is assumed that matter only couples to gravity and not to $\phi$. 

The field $\lambda$ is a Lagrange multiplier enforcing the \emph{mimetic constraint}
\be\label{EQ:MimeticConstraint}
g^{\mu\nu}\pa_\mu\phi\pa_\nu\phi=1 ~.
\ee
Hence, the $f(\chi)$ term in the action (\ref{Action}) does not introduce higher-order derivatives in the equations of motion \cite{Chamseddine:2016uef}.
The mimetic constraint (\ref{EQ:MimeticConstraint}) implies the existence of a privileged ---albeit dynamical--- spacetime foliation, with time function $t=\phi$. 
  In order to show this, it is useful to consider an ADM decomposition of the metric
 \be\label{ADM}
 \de s^2=N^2\de t^2 -h_{ij}(\de x^i+N^i\de t )(\de x^j+N^j\de t ) ~,
 \ee
 where $h_{ij}$ is the (positive definite) spatial metric on $\Sigma_\phi$ and $x^i$ are spatial coordinates. $N$ and $N^i$ denote the lapse function and shift vector, respectively. In this foliation, the mimetic constraint (\ref{EQ:MimeticConstraint}) implies $N=1$, while $N^i$ is undetermined. Let us define the vector field $u^\mu=g^{\mu\nu}\pa_\nu \phi$.
Clearly, $u^\mu$ has unit norm and is orthogonal to the constant-$\phi$ hypersurfaces $\Sigma_\phi$. Moreover, $u^\mu$ generates an irrotational geodesic congruence: the geodesic property is an immediate consequence of the mimetic constraint\footnote{In fact, we have $u^\mu\nabla_\mu u_\nu=u^\mu\nabla_\mu \nabla_\nu\phi=u^\mu\nabla_\nu \nabla_\mu\phi=\frac{1}{2} \nabla_\nu (u^\mu u_\mu)=0$.}  (\ref{EQ:MimeticConstraint}), and one has for the vorticity $\omega_{\mu\nu}=\nabla_{[\nu}u_{\mu]}=\nabla_{[\nu}\nabla_{\mu]}\phi=0$.\footnote{Alternatively, vorticity can be shown to vanish using Frobenius theorem (see e.g.~\cite{Wald:1984rg}).}
  Therefore, it is convenient to make a gauge choice such that the $x^i$ represent comoving coordinates for the geodesic congruence considered; i.e.~we set $N^i=0$ in (\ref{ADM}). With this gauge choice the time-flow vector field is $t^\mu\frac{\pa}{\pa x^\mu}=\frac{\pa}{\pa\phi}$ 
and  
the metric reads as
\be\label{Eq:MetricPreferredFoliation}
\de s^2= \de \phi^2 -h_{ij}\de x^i\de x^j ~.
\ee
Such a foliation is 
referred to as the $\phi$-time gauge.
It is clear from (\ref{Eq:MetricPreferredFoliation}) that the scalar $\phi$ represents proper time as measured by a preferred family of freely-falling observers.

At this stage we would like to give a geometric argument for the absence of higher-order derivatives in the dynamical equations. From the definition of $\chi$, we have
\be
\chi=\Box\phi=\nabla_\mu u^\mu=\theta ~,
\ee
where $\theta$ denotes the expansion of the irrotational geodesic congruence generated by $u^\mu$.\footnote{An extension of mimetic gravity  including vorticity was provided in Ref.~\cite{Barvinsky:2013mea} 
with the introduction of a generalized Proca vector field.} 
 Thus, we recognize that the $f(\chi)$ term in (\ref{Action}) depends only on the expansion scalar as computed in the privileged foliation.\footnote{A class of theories generalizing mimetic gravity and admitting a preferred foliation was considered in Ref.~\cite{deHaro:2017yll}; unlike mimetic gravity though the formulation of these theories is manifestly non-covariant.}  
 This is a crucial property of the theory, which makes singularity resolution possible.

\subsection{Equations of Motion}

We will now derive the remaining equations of motion for the theory considered, following Ref.~\cite{Chamseddine:2016uef}. The gravitational field equations are obtained by varying (\ref{Action}) w.r.t.~the metric
\be\label{EQ:FieldEquations}
G_{\mu\nu}=T_{\mu\nu}^{\psi}+\tilde{T}_{\mu\nu} ~.
\ee
The matter stress-energy tensor is defined as
\be
T_{\mu\nu}^{\psi}=\frac{2}{\sqrt{-g}}\frac{\delta S_{\rm m}}{\delta  g^{\mu\nu}} ~.
\ee
The extra term on the r.h.s.~of Eq.~(\ref{EQ:FieldEquations}) is the stress-energy tensor of the $\phi$-sector
\be\label{EffectiveStressEnergy}
\tilde{T}_{\mu\nu}=2\lambda \pa_\mu\phi\pa_\nu\phi+g_{\mu\nu}(\chi f_{\chi}-f+g^{\rho\sigma}\pa_{\rho}f_{\chi}\pa_\sigma\phi)-(\pa_\mu f_{\chi}\pa_\nu\phi+\pa_\nu f_{\chi}\pa_\mu\phi) ~.
\ee

Varying the action w.r.t.~$\phi$ we obtain
\be\label{EQ:scalarEOM}
\Box f_\chi-2\nabla^{\mu}(\lambda\pa_{\mu}\phi)=0 ~,
\ee
which allows us to eliminate the Lagrange multiplier. We observe that Eq.~(\ref{EQ:scalarEOM}) can be interpreted as a continuity equation; i.e.~one has  $\nabla^\mu J_\mu=0$, where the current $J_\mu$ is defined as
\be
J_\mu=\pa_\mu f_\chi-2\lambda\pa_{\mu}\phi ~.
\ee
In fact, it can be shown that $J_\mu$ is the Noether current corresponding to the invariance of the action under constant shifts of the scalar field $\phi$, i.e.~$\phi\to\phi+\alpha$. Shift-invariance symmetry of mimetic gravity, and the associated Noether current have been studied earlier in Refs.~\cite{Chamseddine:2014vna,Mirzagholi:2014ifa,Hammer:2015pcx} for the particular case~$f(\chi)=\frac{1}{2}\gamma \chi^2$.

Finally, the equations of motion for matter are obtained by varying the action w.r.t.~$\psi$. Clearly, their precise form depends both on the particular type of field $\psi$ considered and on the specific form of the action. Nevertheless, given our assumption that matter only couples to the metric, diffeomorphism invariance of the matter action $S_{\rm m}$ implies that its stress-energy tensor is covariantly conserved \cite{Wald:1984rg}. Similar considerations apply to the $\phi$-sector. Thus, we have
\be\label{EQ:StressEnergyConservation}
\nabla^\mu T^\psi_{\mu\nu}=0~, \hspace{1em} \nabla^\mu \tilde{T}_{\mu\nu}=0 ~.
\ee
We stress that the two conservation laws in (\ref{EQ:StressEnergyConservation}) are independent, due to the absence of direct couplings between $\phi$ and ordinary matter in the action (\ref{Action}).

\subsection{Some Remarks on the Relation with Other Models}
If one were to drop the $f(\chi)$ term from the action (\ref{Action}), the ensuing model would correspond to a particular case of the Brown-Kucha\v{r} action for general relativity minimally coupled to dust~\cite{Brown:1994py}. In Ref.~\cite{Brown:1994py}, dust is described by means of a set of eight scalar fields: four of them are given by the proper-time and comoving coordinates of dust particles; the remaining four correspond to energy density and momentum (see also \cite{Brown:1992kc}). In the case of irrotational dust, the existence of a privileged foliation allows to reduce the number of fields to just two, namely the dust proper time $\phi$ and the energy density $M$. The reduced Brown-Kucha\v{r}  action thus reads
\be\label{Action_BK}
S_{\rm BK}^{\rm\scriptscriptstyle red.}[g_{\mu\nu},\phi,\lambda]=-\frac{1}{2}\int\de^4x \sqrt{-g}\; \Big(R-M(g^{\mu\nu}\pa_\mu\phi\pa_\nu\phi-1)\Big) ~.
\ee

The Brown-Kucha\v{r} action plays an important role in quantum gravity. In fact, it was suggested in Ref.~\cite{Brown:1994py} that the introduction of pressureless dust could provide a solution to the problem of time. This approach has been pursued in the context of loop quantum gravity in Refs.~\cite{Giesel:2007wn,Husain:2011tk} (see \cite{Husain:2011tm} for applications to quantum cosmology). In particular, the action~(\ref{Action_BK}) leads to a 
Hamiltonian constraint which is linear in the canonical momentum of the scalar field $\phi$ \cite{Husain:2011tk}.

A generalization of the reduced Brown-Kucha\v{r} action  sharing some features with mimetic gravity appeared earlier in Ref.~\cite{Lim:2010yk}.
An effective description of mimetic gravity in terms of an imperfect fluid was considered in Ref.~\cite{Mirzagholi:2014ifa} (see also earlier works \cite{Deffayet:2010qz,Pujolas:2011he} for a similar approach). The mimetic constraint (\ref{EQ:MimeticConstraint}) has been also implemented in modified gravity theories, see e.g.~the reviews~\cite{Sebastiani:2016ras,Nojiri:2017ncd}.\footnote{For the Hamiltonian analysis of different mimetic gravity models, including mimetic $f(R)$, see Ref.~\cite{Ganz:2018mqi}.}
Finally, we note that the constraint (\ref{EQ:MimeticConstraint}) is also used in Einstein-aether theory when the aether vector field is restricted to be hypersurface orthogonal \cite{\aether}.

It was shown in Ref.~\cite{Ramazanov:2016xhp} that the action (\ref{Action}) for mimetic gravity with $f(\chi)=\frac{1}{2}\gamma\chi^2$ is equivalent to the IR limit of projectable Ho\v{r}ava-Lifshitz gravity \cite{Blas:2010hb,Horava:2009uw}. This implies, in particular, that the theory has an extra propagating scalar degree of freedom compared to general relativity and the original Brown-Kucha\v{r} action. The extra degree of freedom is also present for a general $f(\chi)$ (provided that $f_{\chi\chi}$ is not identically vanishing), as confirmed by the canonical analysis performed in Ref.~\cite{Bodendorfer:2017bjt}.

\subsection{Cosmological Sector}
We shall now assume the Friedmann-Lema\^itre-Robertson-Walker (FLRW) spacetime geometry
\be\label{EQ:FRLWansatz}
\de s^2=\de t^2-a^2(t) \bar{h}_{ij}\de x^i\de x^j ~,
\ee
where $x^i$ are comoving coordinates, $t$ is proper time as measured by a comoving observer, and $a(t)$ represents the scale factor. We will consider a flat universe $\bar{h}_{ij}=\delta_{ij}$ for simplicity.
The stress-energy tensor of matter $T_{\mu\nu}^{\psi}$ must respect the symmetry of the background geometry~(\ref{EQ:FRLWansatz}). In particular, we will assume that it takes the perfect fluid form
\be\label{EQ:MatterStressEnergy}
T_{\mu\nu}^{\psi}=(\epsilon+P) u_\mu u_\nu -g_{\mu\nu} P~,
\ee
with $\epsilon$ and $P$ depending only on cosmic time $t$. We observe that the fluid four-velocity $u_\mu$ must coincide with the gradient of the scalar field $\phi$ due to the mimetic constraint (\ref{EQ:MimeticConstraint}), which in this case reads as $\dot{\phi}^2=1$. Thus, $\phi$ and $t$ are affinely related, i.e.~$\phi=\phi_0\pm t$, where $\phi_0$ is a constant. Since the theory is shift-invariant, without loss of generality we shall henceforth assume $\phi=t$. 

Similar considerations as above apply to the stress-energy tensor $\tilde{T}_{\mu\nu}$ arising from the $\phi$-sector. In particular, $\tilde{T}_{\mu\nu}$ is diagonal in the coordinate system considered, with components
\begin{align}
\tilde{T}_{00}=2\lambda+\chi f_{\chi}-f- f_{\chi\chi}\dot{\chi} ~, \label{TimeTimeCompEffTmunu}\\
\tilde{T}_{ii}=-a^2(\chi f_{\chi}-f+ f_{\chi\chi}\dot{\chi}) ~.\label{SpaceSpaceCompEffTmunu}
\end{align}
The Lagrange multiplier $\lambda$ can be eliminated by solving Eq.~(\ref{EQ:scalarEOM}) \cite{Chamseddine:2013kea}, which in the cosmological case boils down to
\be
\pa_t\left[a^3 (f_{\chi\chi}\dot{\chi}-2\lambda) \right]=0 ~.
\ee
Solving for $\lambda$, we obtain
\be\label{SolutionLambda}
\lambda=\frac{C}{2a^3}+\frac{1}{2}f_{\chi\chi}\dot{\chi} ~,
\ee
where $C$ is an integration constant with dimensions of energy. Substituting this solution in Eqs.~(\ref{TimeTimeCompEffTmunu}), (\ref{SpaceSpaceCompEffTmunu}) we obtain
\begin{align}
\tilde{T}_{00}&=M+\tilde{\epsilon} ~,  \label{EQ:MimeticStressEnergy00}\\
\tilde{T}_{ii}&=a^2 \tilde{P} 
~. \label{EQ:MimeticStressEnergyii}
\end{align}
where we defined
\begin{align}
&M=\frac{C}{a^3} ~,\label{EQ:MimeticEnergyDensity}\\
&\tilde{\epsilon}=\chi f_{\chi}-f=\chi^2\frac{\de}{\de\chi}\left(\frac{f}{\chi}\right) ~, \label{EQ:EffectiveFluidEnergyDensity}\\
&\tilde{P}=-(\tilde{\epsilon}+ f_{\chi\chi}\dot{\chi}) ~.  \label{EQ:EffectiveFluidPressure}
\end{align}
Thus, we observe that $\tilde{T}_{\mu\nu}$ receives two distinct contributions: pressureless dust (dubbed `mimetic dark matter' \cite{Chamseddine:2013kea}) with energy density $M$, and an `effective fluid'  with energy density~$\tilde{\epsilon}$ and pressure~$\tilde{P}$. Equation~(\ref{EQ:MimeticEnergyDensity}) suggests that the integration constant $C$ shall be interpreted as the total mass of dust particles contained in a comoving volume, which is a conserved quantity.

We observe that the equation of state of the `effective fluid' depends on the specific functional form of $f(\chi)$, as well as on the evolution of $\chi$ itself. In fact, the equation-of-state parameter is
\be\label{EQ:EOS_EffFluid}
\tilde{w}=\frac{\tilde{P}}{\tilde{\epsilon}}=-\left(1+\frac{f_{\chi\chi}}{\tilde{\epsilon}}\dot{\chi}  \right) ~.
\ee
Note that the phantom divide is not crossed, i.e.~$\tilde{w}\geq-1$, as long as the following condition is satisfied
\be\label{EQ:PhantomDivide}
\frac{f_{\chi\chi}}{\tilde{\epsilon}}\dot{\chi}\leq 0 ~.
\ee
The signs of the quantities involved are not known a priori; in particular, $\tilde{\epsilon}$ is not constrained to take positive values. Therefore, in order to verify whether the inequality (\ref{EQ:PhantomDivide}) holds one must first study the dynamics. Thus, a specific form of $f(\chi)$ must be assigned and the l.h.s.~of (\ref{EQ:PhantomDivide}) must be evaluated on a solution.

Let us now turn to the gravitational field equations in the cosmological model considered. Recalling the definition $\chi=\Box\phi$, and given the line element (\ref{EQ:FRLWansatz}), we have
$\chi=3H$~,
where $H=\frac{\dot{a}}{a}$ is the Hubble expansion rate.
In the case at hand, the field equations (\ref{EQ:FieldEquations}) read as
\begin{align}
\frac{1}{3}\chi^2=\epsilon+\tilde{\epsilon}+M ~, \label{EQ:FriedmannMimetic}\\
2\dot{\chi}+\chi^2=-3(P+\tilde{P}) ~. \label{EQ:SecondFriedmannMimetic}
\end{align}
Equation (\ref{EQ:FriedmannMimetic}) represents an effective Friedmann equation for mimetic gravity.
Combining Eqs.~(\ref{EQ:FriedmannMimetic}) and (\ref{EQ:SecondFriedmannMimetic}), we obtain the equation determining the change of $\chi$ over time
\be\label{EQ:DerChi}
\dot{\chi}=-\frac{3}{2}\left[(\epsilon+P)+(\tilde{\epsilon}+\tilde{P})+M \right]=-\frac{3}{2}\left[(1+w)\epsilon+(1+\tilde{w})\tilde{\epsilon}+M \right]
\ee

All matter components, i.e.~ordinary matter, dust, and the effective fluid, satisfy energy conservation
\begin{align}
\dot{\epsilon}+\chi (\epsilon+P)=0 ~,\\
\dot{\tilde{\epsilon}}+\chi (\tilde{\epsilon}+\tilde{P})=0 ~,\\
\dot{M}+\chi M=0 ~.
\end{align}
Note that there is no energy transfer between different components in the homogeneous case, as each of them undergoes adiabatic expansion.

Using Eqs.~(\ref{EQ:EOS_EffFluid}) and (\ref{EQ:DerChi}), the equation of state parameter for the effective fluid is completely determined by the energy density and pressure of ordinary matter, and by the second derivative of~$f(\chi)$
\be\label{EqOfState}
\tilde{w}=-1+\frac{3}{2}\frac{f_{\chi\chi}}{(1-\frac{3}{2}f_{\chi\chi})\tilde{\epsilon}}\left[(1+w)\epsilon+M \right] ~.
\ee
If we assume that both ordinary matter and mimetic dark matter satisfy the null energy condition  \cite{Wald:1984rg,Rubakov:2014jja}\footnote{In the case at hand, the null energy condition is equivalent to the following inequalities: $(1+w )\epsilon\geq0$ and $M\geq0$. In particular, given Eq.~(\ref{EQ:MimeticEnergyDensity}), such condition requires that a positive value must be chosen for the integration constant~$C$.}%
, the condition for not crossing the phantom divide can be expressed as
\be
\frac{f_{\chi\chi}}{(1-\frac{3}{2}f_{\chi\chi})\tilde{\epsilon}}\geq 0 ~.
\ee 
Note that $\tilde{\epsilon}$ does not have a definite sign in principle, since it must be determined from the functional form of $f(\chi)$ using Eq.~(\ref{EQ:EffectiveFluidEnergyDensity}).

The equation of state (\ref{EqOfState}) can be recast in a more convenient form by recalling that the effective speed of sound $c_{s}^2$, as obtained in Ref.~\cite{Firouzjahi:2017txv} for a general $f(\chi)$ (see also Ref.~\cite{Chamseddine:2014vna} for the quadratic case), is given by\footnote{It is worth mentioning that in the original formulation of mimetic gravity (Ref.~\cite{Chamseddine:2013kea}) the speed of sound is exactly vanishing, which motivated the introduction of higher derivative terms in the action in Ref.~\cite{Chamseddine:2014vna}. The sound speed is vanishing also in mimetic Horndeski models \cite{Arroja:2015yvd}.}
\be\label{Eq:SpeedOfSound}
c_{s}^2=\frac{1}{2}\frac{f_{\chi\chi}}{(1-\frac{3}{2}f_{\chi\chi})} ~.
\ee
Equation (\ref{Eq:SpeedOfSound}) gives the propagation speed of scalar perturbations over a FLRW background when the mimetic scalar field $\phi$ accounts for the totality of the energy density of the universe. Thus, setting $\epsilon=w=0$ in Eq.~(\ref{EqOfState}), and using Eq.~(\ref{Eq:SpeedOfSound}), we have for the equation of state of the effective fluid
\be\label{EqOfState2}
\tilde{w}=-1+\frac{3M}{\tilde{\epsilon}} c_{\rm s}^2 ~.
\ee
It is worth remarking that deviations from Eq.~(\ref{EqOfState2}) are in principle possible when a matter sector is included in the action (\ref{Action}).  
Under the assumptions made above, the phantom divide is not crossed (i.e.~$\tilde{w}\geq-1$) if and only if $\tilde{\epsilon}$ and $c_{\rm s}^2$ have the same sign. 
The adiabatic component of the speed of sound of the effective fluid is given by (see e.g.~Refs.~\cite{Bean:2003fb,Hannestad:2005ak})
\be\label{AdiabaticSpeedOfSound}
c^2_{a}=
\frac{\dot{\tilde{P}}}{\dot{\tilde{\epsilon}}}=w-\frac{\dot{\tilde{w}}}{3H(1+\tilde{w})} ~~.
\ee
Using Eqs.~(\ref{EqOfState2}) and (\ref{AdiabaticSpeedOfSound}), we find the following relation between the effective speed of sound and the adiabatic one
\be
c_a^2=\frac{3M}{\tilde{\epsilon}}c_s^2-\frac{1}{\chi}\frac{\de}{\de t}\log\left(\frac{c_s^2}{\tilde{\epsilon}}\right) ~.
\ee
Thus, in general the effective speed of sound does not coincide with its adiabatic value. This is to be ascribed to dissipative processes occurring in the non-homogeneous case, which are responsible for entropy perturbations \cite{Bean:2003fb,Hannestad:2005ak}. Moreover, we observe that the effective speed of sound can be imaginary (i.e.~$c_s^2<0$), depending on the profile of $f(\chi)$; this phenomenon is due to the well-studied gradient instability of 
mimetic gravity \cite{Ijjas:2016pad,Firouzjahi:2017txv,Hirano:2017zox,Takahashi:2017pje}. For $c_s^2>0$ instead the theory has a ghost instability \cite{Firouzjahi:2017txv}. Gradient and ghost instabilities in mimetic gravity were first studied in Ref.~\cite{Ramazanov:2016xhp} for a quadratic $f(\chi)$, exploiting the equivalence with the IR limit of projectable Ho\v{r}ava-Lifshitz gravity.

\section{Relational Evolution in Mimetic Gravity}\label{Sec:Relational}
In this section we recast the evolution equations for the cosmological sector of mimetic gravity in a {\it relational} form. Thus, suitable reference matter fields {\it consistently coupled} to gravity are introduced so as to provide a {\it physically realized} coordinate system.
This approach was pioneered by DeWitt \cite{witten1962gravitation,DeWitt:1967yk} and further developed in Refs.~\cite{Isham:1984sb,Rovelli:1990ph,Kuchar:1990vy,Brown:1994py} (see also \cite{Kuchar:1991qf,Gambini:2004pe} and references therein).
Such a relational reformulation of the dynamics is particularly useful in order to establish a precise connection between mimetic gravity and cosmological models obtained from background-independent approaches to quantum gravity. This will be done in Section~\ref{Sec:GFTbackground}.

We consider a minimally coupled massless scalar $\psi$ playing the role of a matter clock.
The matter Lagrangian is then given by
\be\label{LagrangianPsi}
L_{\rm m}=\frac{1}{2}g^{\mu\nu}\pa_\mu\psi \pa_\nu\psi ~.
\ee
Massless scalars are commonly used as material references in quantum gravity \cite{Giesel:2012rb,Kuchar:1995xn,Rovelli:1993bm,\GFTbounce,Gielen:2018fqv} and quantum cosmology \cite{Blyth:1975is,Smolin:1993ka,Alexander:2003wb,Ashtekar:2011ni}.
The stress-energy tensor obtained from (\ref{LagrangianPsi}) is 
\be\label{EQ:StressEnergyScalar}
T^{\psi}_{\mu\nu}=\pa_\mu\psi \pa_\nu\psi-\frac{1}{2}g_{\mu\nu}(g^{\rho\sigma}\pa_{\rho}\psi \pa_{\sigma}\psi ) ~.
\ee
In the cosmological case $T^{\psi}_{\mu\nu}$ takes the perfect fluid form, Eq.~(\ref{EQ:MatterStressEnergy}), with equation of state parameter~$w=1$ and energy density given by
$\epsilon=\frac{1}{2}\dot{\psi}^2$~.
Extremizing the action (\ref{Action}) w.r.t.~$\psi$ we obtain the Klein-Gordon equation
\be\label{EQ:KleinGordonPsi}
\Box\psi=\frac{1}{\sqrt{-g}}\pa_\mu\left(\sqrt{-g}\,g^{\mu\nu}\pa_\nu \psi\right)=0 ~.
\ee
Equation~(\ref{EQ:KleinGordonPsi}) states that the current $\pa_\mu\psi$ corresponding to $\psi$-shift symmetry is conserved.
Assuming a FLRW cosmological background, with line element (\ref{EQ:FRLWansatz}), Eq.~(\ref{EQ:KleinGordonPsi}) becomes a conservation law for the canonical momentum $p_\psi$ of the scalar field $\psi$~; i.e.~we have $\dot{p}_\psi=0$, with $p_\psi\equiv a^3\dot{\psi}$.
Such a conservation law 
implies in particular that, if $p_\psi\neq0$, $\dot{\psi}$ is nowhere vanishing and keeps the same sign throughout cosmic evolution. Thus, the scalar field $\psi$ is globally monotonic and therefore represents a well-behaved matter clock for the system. From this point of view, the Klein-Gordon equation (\ref{EQ:KleinGordonPsi}) can be regarded as the requirement that $\psi$ be a 
harmonic time coordinate \cite{Kuchar:1991pq,Gielen:2018fqv}. 

The FLRW line element (\ref{EQ:FRLWansatz}) can be re-expressed in  the $\psi$-time gauge  as
\be
\de s^2=N^2(\psi)\, \de\psi^2-a^2(\psi) \de {\bf x}^2 ~.
\ee
The lapse function is given by
\be
N(\psi)=(\dot{\psi})^{-1}=\frac{a^3}{p_\psi} ~.
\ee
We can then define the relational Hubble expansion rate as
$\mathcal{H}=\frac{a^{\prime}}{a}$,
where the prime denotes differentiation w.r.t.~the matter clock $\psi$, i.e.
$a^{\prime}\equiv\frac{\de a}{\de\psi}$~.
Recalling the definition of $\chi$, we have the following relation between $\chi$ and $\mathcal{H}$
\be\label{Eq:RelationChi_H}
\chi=3\,p_\psi \frac{\mathcal{H}}{a^3}~,
\ee
whence
\be\label{EQ:Chi2RelH2}
\frac{1}{3}\chi^2=6\mathcal{H}^2  \epsilon\\~.
\ee
Thus, using (\ref{EQ:Chi2RelH2}) we obtain a relational version of the effective Friedmann equation (\ref{EQ:FriedmannMimetic})
\be\label{EQ:FriedmannRelFirstStep}
\mathcal{H}^2=\frac{1}{6}\left(1+\frac{\tilde{\epsilon}}{\epsilon}+\frac{M}{\epsilon}\right) ~.
\ee
Recalling the definition of the effective fluid energy density, Eq.~(\ref{EQ:EffectiveFluidEnergyDensity}), the relational Friedmann equation (\ref{EQ:FriedmannRelFirstStep})  can be recast as
\be\label{Eq:ODE_for_f}
\mathcal{H}^2\left[1-3 \frac{\de}{\de\chi}\left(\frac{f}{\chi}\right) \right]=\frac{1}{6}\left(1+\frac{M}{\epsilon}\right) ~.
\ee
Equation (\ref{Eq:ODE_for_f}) makes manifest the corrections introduced by the $f(\chi)$ term in (\ref{Action}). Indeed, this will be the form that will be most useful for the applications considered in the remainder of this article. 

Let us now turn to the derivation of the acceleration equation in the relational framework. The following chain of equalities holds
\be\label{EQ:RelationalAcceleration}
\frac{\ddot{a}}{a}=\dot{H}+H^2=\frac{1}{3}\left(\dot{\chi}+\frac{1}{3}\chi^2\right)=2\epsilon\left(\mathcal{H}^{\prime}-2\mathcal{H}^2 \right) ~.
\ee
The r.h.s.~of Eq.~(\ref{EQ:RelationalAcceleration}) thus provides a purely relational definition of the acceleration. It can be verified that such a definition is equivalent to the one given in Refs.~\cite{deCesare:2016axk,deCesare:2016rsf}.

\section{Bouncing Cosmological Backgrounds from Quantum Gravity}\label{Sec:GFTbackground}

Nonsingular bouncing cosmological backgrounds have been recently obtained in the group field theory approach to quantum gravity \cite{\GFTbounce}.
 In this fully background-independent approach, spacetime is emergent from the collective behaviour of a large number of quanta of geometry.\footnote{For a detailed review of the group field theory formalism and the emergent cosmology scenario the reader is referred to Ref.~\cite{Gielen:2016dss}.}
Such quanta can be represented as quantum tetrahedra \cite{Gielen:2016dss} carrying algebraic data (i.e.~spin labels) that characterize their geometry, namely~the area of their faces and their volume.\footnote{Area and volume are kinematical geometric operators, see e.g.~\cite{Thiemann:2002nj}. Their definition in group field theory is inherited from loop quantum gravity, since the former provides a second quantized reformulation of the latter, see Ref.~\cite{Oriti:2013} for details. The spin labels mentioned above characterize the spectrum of such operators.}
Homogeneous cosmology is then recovered by considering a particular class of states (i.e.~condensates \cite{Gielen:2013naa}) in the quantum theory, and studying their mean-field dynamics.\footnote{Some analogies with the phenomenon of Bose-Einstein condensation have been suggested in Ref.~\cite{Oriti:2016acw}.}
The evolution of geometric observables (such as the volume) must be defined with respect to other dynamical fields. In particular, a massless scalar field can be used as a matter clock \cite{\GFTbounce}.\footnote{See Ref.~\cite{Li:2017uao} for a more general construction involving multiple scalar fields.}

The effective Friedmann dynamics obtained in Ref.~\cite{\GFTbounce} describes the evolution of the emergent cosmological background in relational terms. Within the class of quantum states considered in \cite{\GFTbounce}, the simplest condensates are such that all tetrahedra are isotropic and have the same volume (single-spin states). In this particular case, the effective Friedmann equation takes a remarkably simple form  
\be\label{Eq:RelationalFriedmann}
\mathcal{H}^2=\frac{4}{9}\left(3\pi G+\frac{E v_o}{V}-\frac{Q^2 v_o^2}{V^2}\right) ~,
\ee
where Newton's constant has been reinstated. In Eq.~(\ref{Eq:RelationalFriedmann}) $v_o$ denotes the volume of a quantum tetrahedron in the condensate, $V$ is the total volume of the system, and $\mathcal{H}$ is the relational Hubble rate. The scale factor can be defined as $a\propto V^{1/3}$. We remark that in this approach the scale factor is a derived quantity, whereas primary physical quantities (e.g.~the volume) are computed as expectation values of geometric observables.
$E$ and $Q$ are conserved quantities.\footnote{In particular,  it was shown to in Ref.~\cite{\GFTbounce} that $Q$ should be identified with the canonical momentum of the scalar field. Nevertheless, we shall relax this condition here so as to keep our analysis as general as possible, compatibly with the structure of the background equation (\ref{Eq:RelationalFriedmann}). By doing so, we encompass both the background dynamics obtained from the full theory in Ref.~\cite{\GFTbounce} and the one obtained in the toy model of  Ref.~\cite{Adjei:2017bfm}, which differ by the values of the parameters $E$, $Q$.
The relational Hamiltonian model considered in Ref.~\cite{Wilson-Ewing:2018mrp}, in the case of a single dominant spin component, yields the background dynamics (\ref{Eq:RelationalFriedmann}) with $E=0$.\\~\\}  
The last two terms in Eq.~(\ref{Eq:RelationalFriedmann}) represent quantum corrections to the standard Friedmann dynamics.

It is convenient to rewrite Eq.~(\ref{Eq:RelationalFriedmann}) as
\be\label{EQ:RelationalFriedmann_alphabeta}
\mathcal{H}^2=\frac{1}{6}+\frac{\alpha}{V}-\frac{\beta}{V^2} ~,
\ee
having defined $\alpha\equiv\frac{4}{9}E v_o$, $\beta\equiv\frac{4}{9}Q^2 v_o^2$, and recalling our choice of units $8\pi G=1$. Since $\mathcal{H}^2\geq0$, the r.h.s.~of Eq.~(\ref{EQ:RelationalFriedmann_alphabeta}) must be non-negative.
When $\mathcal{H}$ vanishes, the universe undergoes a bounce and the volume attains its minimum value
\be\label{EQ:Vmin}
V_{\rm min}=-3\alpha+\sqrt{9\alpha^2+6\beta} ~.
\ee
Note that $V_{\rm min}$ is strictly positive. The bounce is generic for $\beta>0$ and for all values of $\alpha$.

Lastly, we note that switching to proper time gauge using Eq.~(\ref{EQ:Chi2RelH2}), and recalling $\epsilon=p_\psi^2/(2V^2)$, the effective Friedmann equation (\ref{EQ:RelationalFriedmann_alphabeta})  can be recast as 
\be\label{Eq:ChiOfV}
\frac{1}{3}\chi^2=\frac{p_\psi^2}{2 V^2}\left[1+6\left(\frac{\alpha}{V}-\frac{\beta}{V^2}\right)\right]~.
\ee

\section{Reproducing the group field theory background dynamics in mimetic gravity}\label{Sec:GFT}

Our purpose in this section is to reconstruct the functional form of $f(\chi)$ in the action (\ref{Action}) for limiting curvature mimetic gravity, assuming that the background dynamics is determined by the effective Friedmann equation (\ref{EQ:RelationalFriedmann_alphabeta}) obtained in group field theory.
We will assume for definiteness and without loss of generality that $p_\psi>0$, so that the signs of $\chi$ and $\mathcal{H}$ agree (see Eq.~(\ref{Eq:RelationChi_H})). In the effective Friedmann dynamics (\ref{EQ:RelationalFriedmann_alphabeta}) the r.h.s.~receives contributions from the energy density of the scalar field $\psi$ and  quantum gravity corrections represented by the remaining two terms. 
For simplicity, other matter species are not included. In particular, the contribution of dust to the energy density is negligible.
This is a reasonable approximation in the early universe: the scalar field $\psi$ dominates the energy density in such regime since $\epsilon\sim a^{-6}$, while $M\sim a^{-3}$.
Therefore, we have from Eq.~(\ref{Eq:ODE_for_f})
\be\label{Eq:ODE_for_f2}
\mathcal{H}^2\left[1-3 \left(\frac{f}{\chi}\right)_\chi \right]=\frac{1}{6} ~.
\ee

\begin{figure}
\includegraphics[width=0.5\columnwidth]{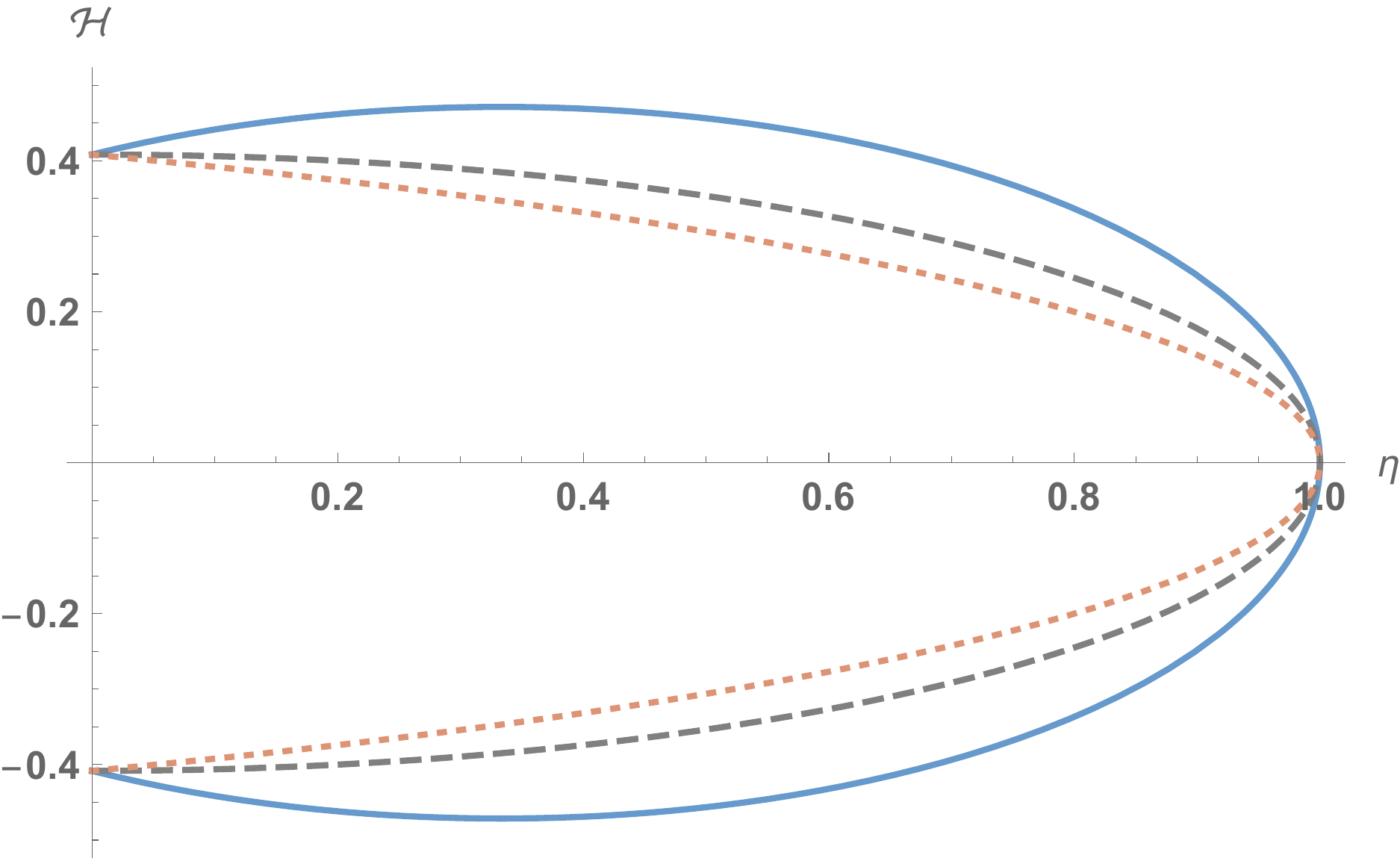}
\caption{Plot of the relational Hubble rate $\mathcal{H}$ as a (multi-valued) function of the inverse volume $\eta$, as obtained from (\ref{EQ:RelationalFriedmann_alphabeta}). The curve is the portion of the ellipse with equation $\mathcal{H}^2=\frac{1}{6}+\alpha \eta - \beta \eta^2$ lying in the half-plane $\eta\geq 0$. The positive (negative) branch corresponds to the expanding (resp.~contracting) phase. The bounce is represented by the vertex of the ellipse on the positive $\eta$ axis. In the plot, the value of the parameter $\alpha$ is chosen so as to have $V_{\rm min}=1$, i.e. $\alpha=\beta-\frac{1}{6}$. The curves in Fig.~\ref{HvsEta} thus only differ by the value of $\beta$: the thick blue curve corresponds to $\beta=\frac{1}{2}$, the dashed gray one to $\beta=\frac{1}{6}$, and the dotted orange one to $\beta=\frac{1}{24}$. The dashed curve corresponds to the LQC case ($\alpha=0$), which separates the curves in two classes. In fact, for $\alpha<0$ both branches of $\mathcal{H}(\eta)$ are monotonic, whereas for $\alpha>0$ they have an extremum for $\eta>0$. We observe that the interesection between the curve and the $\mathcal{H}$ axis does not depend on the value of the parameters; in fact, it is entirely determined by the scalar field contribution (first term) in Eq.~(\ref{EQ:RelationalFriedmann_alphabeta}) and by the choice of units.}\label{HvsEta}
\end{figure}

Equation~(\ref{Eq:ODE_for_f2}) can be regarded as a differential equation for $f(\chi)$, with $\mathcal{H}^2$ satisfying (\ref{EQ:RelationalFriedmann_alphabeta}). Its solution is given by the following integral
\be
f(\chi)=\chi\int\de\chi\; \left(\frac{ \mathcal{H}^2-\frac{1}{6}}{3 \mathcal{H}^{2}}\right) ~,
\ee
where $\mathcal{H}$ must be regarded as a function of $\chi$ through $V$.
The integral can be evaluated by changing the integration variable $\chi=\chi(V)$ and integrating over the volume $V$. Such an operation is allowed in each domain where $\chi$ is a monotonic function of the volume $V$; in particular, it is required that the first derivative $\chi_V$ be non-vanishing and finite.
One finds, in any such domains
\be\label{EQ:SolutionMimetic}
f(\chi)=c\,\chi+\epsilon(\chi)+\frac{1}{3}\chi^2+\frac{p_\psi}{3\sqrt{\beta}}\,\chi \arctan \left(\frac{1}{\sqrt{\beta}}\frac{\de \mathcal{H}}{\de \eta} \right)  ~,
\ee
where $\eta\equiv V^{-1}$ and $c$ is an integration constant. The graph of $ \mathcal{H}(\eta)$ is shown in Fig.~\ref{HvsEta}. It is important to remark that the value of the integration constant $c$ must be specified in each domain where $\chi(V)$ is invertible: suitable matching conditions must then be imposed on $f$ and $f_\chi$; this problem will be addressed in Section \ref{Sec:Matching}.

\begin{figure}
\includegraphics[width=0.5\columnwidth]{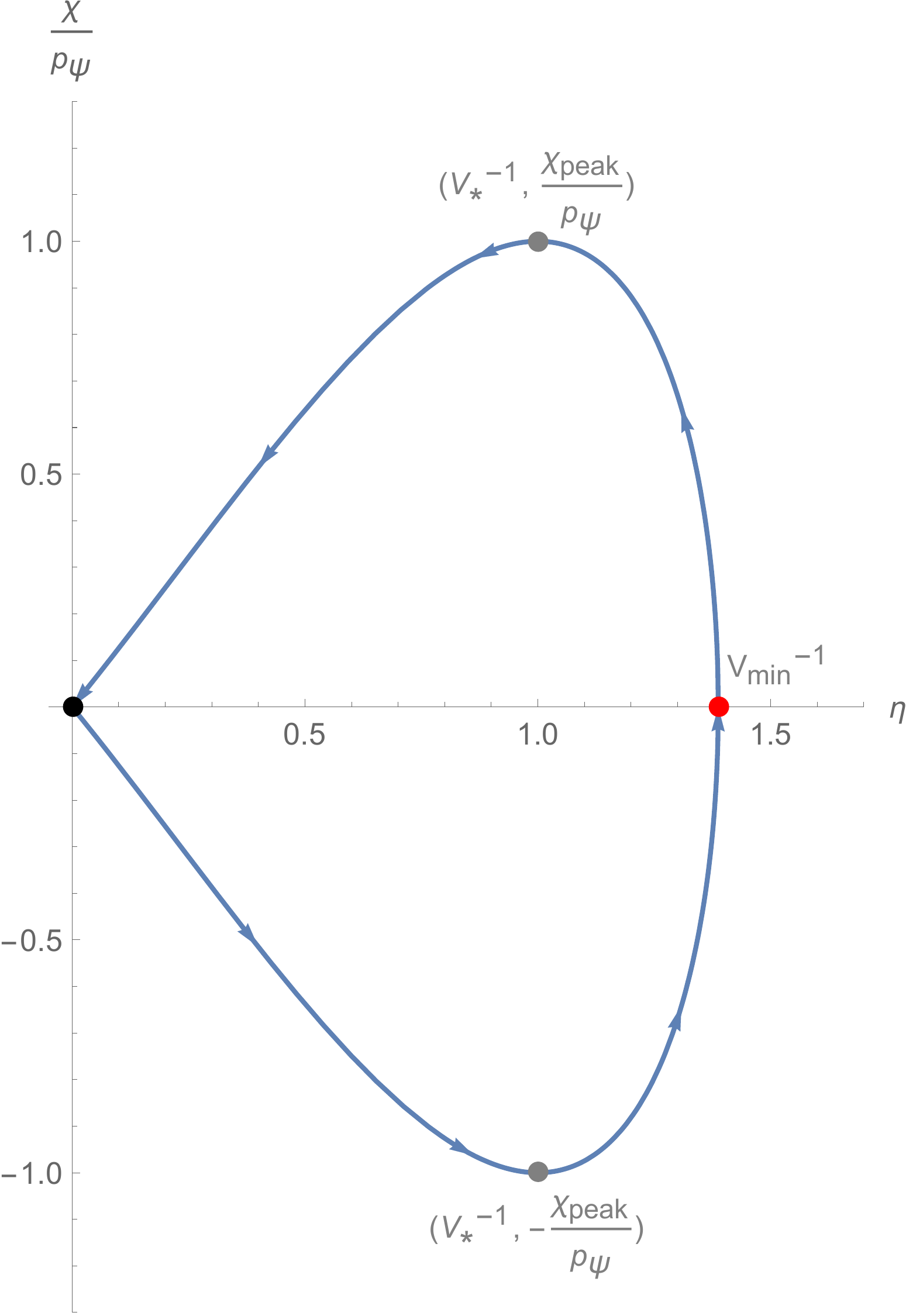}
\caption{The figure shows the graph of the multivalued function $\chi(\eta)$. The contracting branch is contained in the lower-half plane ($\chi<0$), whereas the expanding branch lies in the upper half-plane ($\chi>0$). The two branches meet at the bounce, denoted by the red point. 
The arrows point in the direction of cosmic time flow. We note that the origin of the $(\eta,\chi)$ plane, here denoted by a black point, corresponds to infinite volume and vanishing Hubble rate ($\propto\chi$); it is both the starting point of the contracting branch and the end point of the expanding one. The values of the parameters are chosen so as to have $\chi_{\rm peak}/p_\psi=V_*=1$, i.e.~$\alpha=\frac{1}{9}$, $\beta=\frac{1}{6}$. The qualitative behaviour of the curve is nevertheless generic.}\label{ChiVSEta}
\end{figure}

In order to exhibit explicitly the functional form of $f(\chi)$ one must first invert the function $V=V(\chi)$, and therefore express $\mathcal{H}$ and $\epsilon$ as functions of $\chi$. The inverse function is determined by solving Eq.~(\ref{Eq:ChiOfV}) for $V$, which leads to finding the roots of the following quartic polynomial
\be
P_4(V)\equiv \left(\frac{\chi^2}{9p^2}\right) V^4-\frac{1}{6}\,V^2-\alpha\, V+\beta \label{EQ:QuarticPoly} ~.
\ee
Clearly we must only retain non-negative roots since $V\geq0$. Note that the equal sign can only be achieved in the $\beta=0$ case, as discussed in \cite{deCesare:2016axk}.
We remark that the sign of $\chi$ is negative in the contracting phase, while it is positive in the expanding one. Therefore, in the expanding phase ($\chi>0$) there are two branches
\begin{align}
V^{\rm e}_{\plus}&=\frac{ \sqrt{\sigma}+\sqrt{3-\sigma+\frac{18\alpha}{\sqrt{\sigma}}\frac{\chi}{p_\psi} }}{2\frac{\chi}{p_\psi}} ~, \label{Branch_e+}\\
V^{\rm e}_{\minus}&=\frac{\sqrt{\sigma}-\sqrt{3-\sigma+\frac{18\alpha}{\sqrt{\sigma}}\frac{\chi}{p_\psi} }}{2\frac{\chi}{p_\psi}} ~, \label{Branch_e-}
\end{align}
where $\sigma$ is a function of $\chi$, defined as
\begin{align}
\sigma&=1+\frac{A}{2(-B+Q)^{1/3}}+\frac{(-B+Q)^{1/3}}{2}\\
A&=1+48\beta\frac{\chi^2}{p_\psi^2}\\
B&=1-36(9\alpha^2+4\beta)\frac{\chi^2}{p_\psi^2}\\
Q&=\sqrt{-A^3+B^2}
\end{align}
 The solutions describing the contracting phase ($\chi<0$) are obtained from the above under the transformation $\chi\to-\chi$, and are thus given by
 \begin{align}
V^{\rm c}_{\plus}&=-\frac{ \sqrt{\sigma}+\sqrt{3-\sigma-\frac{18\alpha}{\sqrt{\sigma}}\frac{\chi}{p_\psi} }}{2\frac{\chi}{p_\psi}} ~, \label{Branch_c+}\\
V^{\rm c}_{\minus}&=-\frac{\sqrt{\sigma}-\sqrt{3-\sigma-\frac{18\alpha}{\sqrt{\sigma}}\frac{\chi}{p_\psi} }}{2\frac{\chi}{p_\psi}} ~. \label{Branch_c-}
\end{align}
Solutions (\ref{Branch_c+}), (\ref {Branch_e+}) (denoted by a `$\scriptstyle +$') correspond to the large volume branches, where $|\chi(V)|$ is monotonically decreasing. Solutions (\ref{Branch_c-}), (\ref {Branch_e-}) (denoted by a `$\scriptstyle -$') correspond to the small-volume branches around the bounce where $|\chi(V)|$ is monotonically increasing. Therefore, there are three branching points at finite volumes: the first one is at the bounce, where $\chi=0$ and the volume attains its minimum $V_{\rm min}$; the remaining two branching points correspond to the peaks of $|\chi|$ in the pre- and post-bounce phases. The four different solutions are represented in Fig.~\ref{ChiVSEta} as portions of a curve in the $(\eta,\chi)$ plane, smoothly joined at the branching points. 
In the case $\alpha=0$, the expressions above simplify considerably and one recovers the LQC case.

Given the symmetry of the problem under $\chi\to-\chi$ (time reversal), it is convenient to describe the branches pairwise. More precisely, the two branches around the bounce can be captured at once by
\be
V_{\minus}=\frac{\sqrt{\sigma}-\sqrt{3-\sigma+\frac{18\alpha}{\sqrt{\sigma}}\frac{|\chi|}{p_\psi} }}{2\frac{|\chi|}{p_\psi}} ~. \label{Branch_Bounce}
\ee
Similarly, the two large-volume branches (i.e.~those away from the bounce) are given by
\be
V_{\plus}=\frac{\sqrt{\sigma}+\sqrt{3-\sigma+\frac{18\alpha}{\sqrt{\sigma}}\frac{|\chi|}{p_\psi} }}{2\frac{|\chi|}{p_\psi}} ~. \label{Branch_LargeVol}
\ee

\section{Asymptotics at the Branching Points}\label{Sec:Asymptotics}
The exact functional form of $V(\chi)$ in its different branches is expressed by formulae (\ref{Branch_Bounce}), (\ref{Branch_LargeVol}), whereby the function $f(\chi)$ can be computed exactly using Eq.~(\ref{EQ:SolutionMimetic}). Note that $f(\chi)$ is itself a multi-valued function as a consequence of the multi-valuedness of $V(\chi)$. To be specific, $f(\chi)$ has one branch for each region where $\chi(V)$ is invertible, see Fig.~\ref{ChiVSEta}. 

Although the results obtained in the previous section allow us to compute $f(\chi)$ exactly, it is nevertheless useful to derive more manageable expressions by means of suitable approximations in regimes of physical interest, particularly around the branching points. This is particularly helpful in order to gain better qualitative understanding in the behaviour of the solution, and is also necessary for the matching conditions studied in Section~\ref{Sec:Matching}. 

\subsection{Bounce}
We start focusing on the region around the bounce at $(\chi,V)=(0,V_{\rm min})$. The corresponding branch of $V(\chi)$ can be obtained by Taylor expanding (\ref{Branch_Bounce}). Alternatively, we can find the roots of $P_4(V)$ (defined in Eq.~(\ref{EQ:QuarticPoly})) perturbatively; the results obtained by either method are clearly the same. Thus, we express the volume as its minimum plus a perturbation $V\simeq V_{\rm min}+\delta V$, where $\delta V$ is of the same order as $\chi^2$. Plugging such a perturbative ansatz in the equation $P_4(V)=0$, to zero-th order we determine 
$V_{\rm min}$ as given by Eq.~(\ref{EQ:Vmin}). To first perturbative order, we obtain
\be
\left(\frac{\chi^2}{9p_\psi^2}\right) V_{\rm min}^4-\left(\frac{1}{3}\,V_{\rm min}+\alpha\right) \delta V=0 ~.
\ee
Thus, to second order in $\chi$, we have
\be\label{EQ:VolumeAroundBounce}
V(\chi)= V_{\rm min}\left(1+\frac{V_{\rm min}^3}{V_{\rm min}+3\alpha}\,\frac{\chi^2}{3p_\psi^2}\right) +\mathcal{O}(\chi^4) ~.
\ee
The $\mathcal{O}(\chi^4)$ term can be computed by going to second order in the perturbative expansion; however, it will not be necessary here. The absence of the $\mathcal{O}(\chi^3)$ term is due to symmetry under $\chi\to-\chi$.
Using (\ref{EQ:VolumeAroundBounce}), the energy density can then be approximated as
\be\label{EQ:DensityAroundBounce}
\epsilon(\chi)=\frac{p_\psi^2}{2V_{\rm min}^2}\left(1-2\frac{V_{\rm min}^3}{V_{\rm min}+3\alpha}\,\frac{\chi^2}{3p_\psi^2}\right) +\mathcal{O}(\chi^4)~.
\ee
We now turn to the last term in Eq.~(\ref{EQ:SolutionMimetic}). To begin with, we note that $\chi$ and $\mathcal{H}$ always have the same sign\footnote{Recall that we assumed $p_\psi>0$. See discussion at the beginning of Section~\ref{Sec:GFT}.}; hence, they can be replaced by their absolute values.
\be
\chi \arctan \left(\frac{1}{\sqrt{\beta}}\frac{\de \mathcal{H}}{\de \eta} \right) =|\chi| \arctan \left(\frac{1}{\sqrt{\beta}}\frac{\de |\mathcal{H}|}{\de \eta} \right) ~.
\ee
The derivative of $|\mathcal{H}|$ w.r.t.~the inverse volume $\eta$ is given by
\be
\frac{\de|\mathcal{H}|}{\de \eta}=\frac{\alpha-2\beta \eta}{2|\mathcal{H}|} ~.
\ee
Note that this expression diverges at the bounce, where $\eta=\eta_{\rm max}=V_{\rm min}^{-1}$ (see also Fig.~\ref{HvsEta}). In fact, we have
\be
\lim_{\eta\to\eta_{\rm max}} \frac{\de|\mathcal{H}|}{\de \eta}=-\infty ~,
\ee
since the numerator is negative definite in a neighbourhood of the bounce, while the denominator stays positive since 
\be
\alpha-2\beta \eta_{\rm max}=-\sqrt{\alpha^2+2\beta/3}<0 ~.
\ee
Next, recalling the asymptotic expansion of the arctangent
\be
\lim_{z\to-\infty}\arctan{z}=-\frac{\pi}{2}-\frac{1}{z}+\mathcal{O}\left(\frac{1}{z^3}\right)~,
\ee
and since 
\be
\left(\frac{\de|\mathcal{H}|}{\de \eta}\right)^{-1}=-V^{-2}\left(\frac{\de|\mathcal{H}|}{\de V}\right)^{-1}=-2\left(\frac{V_{\rm min}}{V_{\rm min}+3\alpha}\right)\frac{|\chi|}{p_\psi}+\mathcal{O}(\chi^3) ~,
\ee
we can write
\be\label{EQ:ArcTanAroundBounce}
\frac{p_\psi}{3\sqrt{\beta}}|\chi| \arctan \left(\frac{1}{\sqrt{\beta}}\frac{\de |\mathcal{H}|}{\de \eta} \right)= -\frac{\pi p_\psi}{6\sqrt{\beta}}|\chi|+\frac{2}{3}\left(\frac{V_{\rm min}}{V_{\rm min}+3\alpha}\right)\chi^2+\mathcal{O}(\chi^4) ~.
\ee
Finally, the expansion of $f(\chi)$ around the bounce can be obtained using Eqs.~(\ref{EQ:SolutionMimetic}), (\ref{EQ:DensityAroundBounce}), (\ref{EQ:ArcTanAroundBounce}); it reads as
\be\label{EQ:AsymptoticExpansion_Bounce}
f_{\minus}^{\scriptscriptstyle\rm c,e}(\chi)=c_{\minus}^{\scriptscriptstyle\rm c,e}\,\chi+\frac{p_\psi^2}{2V_{\rm min}^2}+\frac{1}{3}\left(\frac{2 V_{\rm min}+3\alpha}{V_{\rm min}+3\alpha}\right)\chi^2-\frac{\pi p_\psi}{6\sqrt{\beta}}|\chi|+\mathcal{O}(\chi^4) ~.
\ee
We denote by $f_{\minus}^{\scriptscriptstyle\rm c}$ the branch of $f(\chi)$ corresponding to the small volume, contracting phase, whereas $f_{\minus}^{\scriptscriptstyle\rm e}$ corresponds to the small volume, expanding phase. The two branches are characterised by different values of the integration constant $c$, which must be fixed so as to ensure a smooth matching, see Section~\ref{Sec:Matching}. In particular, suitable matching conditions will ensure the cancellation of the non-analytic term $\propto |\chi|$ in Eq.~(\ref{EQ:AsymptoticExpansion_Bounce}), thus showing that $f(\chi)$ is analytic at the bounce.

\subsection{Peak Expansion Rate}
We now look for an approximation of $f(\chi)$ around the two peaks corresponding to maximum expansion rate $|\chi|$. The critical point of $\chi(V)$ (see (\ref{Eq:ChiOfV})) is
\be\label{EQ:Vpeak}
V_*=\frac{1}{2} \left( \sqrt{81 \alpha ^2+48 \beta }-9\alpha \right) ~.
\ee
The corresponding peak values of $\chi$ in the expanding and the contracting branch are given by
\be\label{EQ:DefChiPeak}
\chi(V_*)=\pm\,\chi_{\rm peak} ~~,\hspace{1.5em} \chi_{\rm peak}\equiv\frac{p_\psi}{2 V_*}  \sqrt{3+\frac{9 \alpha }{V_*}}~.
\ee
The parameter $\beta$ has been eliminated by solving Eq.~(\ref{EQ:Vpeak}) for $\beta$, whereby we obtain
\be\label{EQ:eliminateBeta_Peak}
\beta = \frac{V_*}{12} \left(V_*+9 \alpha  \right) ~.
\ee

For definiteness, let us focus now on the expanding branch $\chi>0$. The discussion of the contracting branch proceeds in an analogous fashion.
It is clear that $\chi(V)$ is not invertible in any neighbourhood of $V_*$, since the latter is a critical point. Thus, the inverse function $V^{\rm e}(\chi)$ is multi-valued and has two branches joining at $\chi_{\rm peak}$. Since $V_*$ is a critical point of $\chi(V)$, both branches of its inverse must be non-analytic at $\chi_{\rm peak}=\chi(V_*)$. We will now derive the asymptotic expansion of $V^{\rm e}(\chi)$ around $\chi_{\rm peak}$. We start by considering the Taylor expansion of $\chi^2(V)$ around its maximum
\be\label{EQ:TaylorExpansion_ChiAtPeak}
\chi^2(V)=\chi_{\rm peak}^2 \left(1-2\frac{  (2
   V_*+9 \alpha )}{V_*^2 (V_*+3 \alpha)}(V-V_*)^2+12\frac{ (V_*+5 \alpha )}{V_*^3(V_*+3 \alpha )}(V-V_*)^3\right) +\mathcal{O}\left(V-V_*\right)^4 ~.
\ee
Inverting Eq.~(\ref{EQ:TaylorExpansion_ChiAtPeak}), we obtain the following asymptotic expansion
\be\label{EQ:VolumeAsymptotic}
\begin{split}
V^{\rm e}_{\scrscr\pm}(\chi)=V_*\left(1\pm\sqrt{\frac{  (V_*+3 \alpha )}{2 (2V_*+9 \alpha)}}\sqrt{1-\frac{\chi^2}{\chi_{\rm peak}^2}}+\frac{3}{2}\frac{ (V_*+3 \alpha ) (V_*+5  \alpha)}{ (2 V_*+9 \alpha )^2}\left(1-\frac{\chi^2}{\chi_{\rm peak}^2}\right)\;\right)+\\
+\mathcal{O}\left(1-\frac{\chi^2}{\chi_{\rm peak}^2}\right)^{3/2} ~.
\end{split}
\ee
The same result can also be worked out from the exact functional form of $V^{\rm e}_{\pm}(\chi)$ given in Eqs.~(\ref{Branch_e+}),~(\ref{Branch_e-}).

As a consequence of the branching of $V^{\rm e}(\chi)$ at $\chi_{\rm peak}$, the function $f(\chi)$ will also have two branches in the expanding phase. Their asymptotic behaviour near $\chi_{\rm peak}$ can be obtained from the exact expression of $f(\chi)$ given in Eq.~(\ref{EQ:SolutionMimetic}) by plugging in the asymptotic expansion of the volume,~Eq.~(\ref{EQ:VolumeAsymptotic}). The energy density can be approximated as
\be
\begin{split}
\epsilon(\chi)=\frac{p_\psi^2}{2 V_*^2}\left(1\mp \sqrt{\frac{  (2V_*+6 \alpha )}{(2V_*+9 \alpha)}}\sqrt{1-\frac{\chi^2}{\chi_{\rm peak}^2}}-\frac{3}{2}\frac{  \alpha  ( V_*+3 \alpha )}{
   (2 V_*+9 \alpha )^2}\left(1-\frac{\chi^2}{\chi_{\rm peak}^2}\right)\; \right)+\\
   +\mathcal{O}\left(1-\frac{\chi^2}{\chi_{\rm peak}^2}\right)^{3/2} ~.
\end{split}
\ee
The argument of the arctangent in the third term of (\ref{EQ:SolutionMimetic}) is regular at $\chi_{\rm peak}$; its expansion reads as\footnote{It may be worth noting that the maximum of $\mathcal{H}$ does not coincide with the maximum of $\chi$ for any values of the parameters.}
\be\label{Eq:ExpandArgumentArcTan_atPeak}
\begin{split}
\frac{\de \mathcal{H}}{\de \eta}=&-\frac{\sqrt{V_* (V_*+3 \alpha )}}{2 \sqrt{3}}\left(1\mp \frac{\sqrt{2} (V_*+6 \alpha ) }{ \sqrt{(V_*+3\alpha) (2 V_*+9 \alpha)}}\sqrt{1-\frac{\chi ^2}{\chi_{\rm peak}^2} } + \right. \\
& \left. +\frac{ ( V_*+6 \alpha) (4 V_*+15 \alpha )}{2 (2 V_*+9 \alpha)^2} \left(1-\frac{\chi ^2}{\chi_{\rm peak}^2} \right)\; \right)
+\mathcal{O}\left(1-\frac{\chi ^2}{\chi_{\rm peak}^2} \right)^{3/2}~.
\end{split}
\ee
We can compute the asymptotic expansion of the last term in (\ref{EQ:SolutionMimetic}) using Eq.~(\ref{Eq:ExpandArgumentArcTan_atPeak}) and expanding the arctangent; after eliminating $\beta$ using Eq.~(\ref{EQ:eliminateBeta_Peak}) we obtain
\be
\begin{split}
\arctan \left(\frac{1}{\sqrt{\beta}}\frac{\de \mathcal{H}}{\de \eta} \right)=&-\arctan\left(\sqrt{\frac{V_*+3\alpha}{V_*+9\alpha}}\right)\pm  \sqrt{\frac{V_*+9 \alpha}{4   V_*+18 \alpha }}\sqrt{1-\frac{\chi ^2}{\chi_{\rm peak}^2} }+\\
&+\frac{3}{4}\frac{ \alpha   \sqrt{(V_*+3 \alpha) (V_*+9 \alpha )}}{ (2V_*+9 \alpha)^2}\left(1-\frac{\chi ^2}{\chi_{\rm peak}^2} \right)+\mathcal{O}\left(1-\frac{\chi ^2}{\chi_{\rm peak}^2} \right)^{3/2} ~.
\end{split}
\ee
Therefore, in the regime considered here, the leading order term in the asymptotic expansion of $f(\chi)$ reads as
\be\label{EQ:AsymptoticExpansion_Peak_e}
f^{\rm e}_{\scrscr \pm}(\chi)=c^{\rm e}_{\scrscr \pm}\,\chi+\frac{1}{3}\chi^2 +
  \frac{p_\psi^2}{2V_*^2}\left(1-2u+u\left(1-\frac{\chi ^2}{\chi_{\rm peak}^2} \right) \right)+\mathcal{O}\left(1-\frac{\chi ^2}{\chi_{\rm peak}^2} \right)^{3/2} ~,
\ee
where we defined
\be
u=\sqrt{\frac{V_*+3 \alpha}{V_*+9 \alpha}}\arctan\left(\sqrt{\frac{V_*+3 \alpha}{V_*+9 \alpha}}\right) ~.
\ee
We note that the leading order terms in (\ref{EQ:AsymptoticExpansion_Peak_e}) ---disregarding the linear one, which depends on the integration constant--- do not exhibit dependence on the branch. In fact, it can be verified that such dependence only shows up starting from the terms $\mathcal{O}\left(1-\frac{\chi ^2}{\chi_{\rm peak}^2} \right)^{3/2}$.
This implies in particular that the difference between the limits of the two branches $f^{\rm e}_{\scrscr +}(\chi)$, $f^{\rm e}_{\scrscr -}(\chi)$ at $\chi_{\rm peak}$ is only due to the linear term in (\ref{EQ:AsymptoticExpansion_Peak_e}). In fact, the limit $\chi\to\chi_{\rm peak}$ gives
\be\label{LimitAtPeak}
\lim_{\chi\to\chi_{\rm peak}}f^{\rm e}_{\scrscr \pm}(\chi)=c^{\rm e}_{\scrscr \pm}\,\chi_{\rm peak}+\frac{p_\psi^2}{2V_*^2}\left(\frac{3}{2}\frac{V_*+\alpha}{V_*}-2u\right) ~.
\ee

The contracting branch ($\chi<0$) can be studied following analogous steps as above. Here too there are two branches, namely $f^{\rm c}_{\scrscr +}(\chi)$, $f^{\rm c}_{\scrscr -}(\chi)$. Given the symmetry of the problem, 
all of the terms in (\ref{EQ:SolutionMimetic}) ---~with the exception of the linear one~--- must be the same in the two branches. Thus, around $\chi=-\chi_{\rm peak}$ the following asymptotic expansion holds
\be
f^{\rm c}_{\scrscr \pm}(\chi)=c^{\rm c}_{\scrscr \pm}\,\chi+\frac{1}{3}\chi^2 +
  \frac{p_\psi^2}{2V_*^2}\left(1-2u+u\left(1-\frac{\chi ^2}{\chi_{\rm peak}^2} \right) \right)+\mathcal{O}\left(1-\frac{\chi ^2}{\chi_{\rm peak}^2} \right)^{3/2} ~,
\ee
whence it follows 
\be
\lim_{\chi\to-\chi_{\rm peak}}f^{\rm c}_{\scrscr \pm}(\chi)=-c^{\rm c}_{\scrscr \pm}\,\chi_{\rm peak}+\frac{p_\psi^2}{2V_*^2}\left(\frac{3}{2}\frac{V_*+\alpha}{V_*}-2u\right) ~.
\ee

\section{Matching Conditions}\label{Sec:Matching}
 The function $f(\chi)$ has four branches, and to each of those corresponds an integration constant. There are three branching points, namely at the bounce and at $\chi=\pm\chi_{\rm peak}$.
 The integration constants can be determined by requiring continuity of $f$ and its first derivative $f_{\chi}$ along the curve with equation given by~(\ref{Eq:ChiOfV}), so as to ensure regularity of the energy density $\tilde{\epsilon}$ of the effective fluid, Eq.~(\ref{EQ:EffectiveFluidEnergyDensity}), throughout cosmic evolution. This is equivalent to the continuity of such functions along the curve in Fig.~\ref{ChiVSEta}, with the origin excluded.
Therefore, we shall require that the limit of $f$ as a branching point is approached must not depend on the branch. As shown in the following this leads to three algebraic conditions, thus leaving one integration constant undetermined.\footnote{Note that in a closed universe there would be an extra matching condition at recollapse.} In fact, globally, $f(\chi)$ itself is determined only up to a linear term in $\chi$, since $\chi=\nabla_\mu u^\mu$ and the action (\ref{Action}) is defined up to boundary terms. We can fix such underdeterminacy by further requiring that $f(\chi)$ respects the symmetry of the dynamics under  $\chi\to-\chi$, i.e.~we demand that $f(\chi)$ be even
 \be
 f(\chi)=f(-\chi) ~.
 \ee
 Therefore, the integration constants must obey the following relation
\be\label{EQ:ConditionFromParity}
c^{\rm c}_{\scrscr \pm}=-c^{\rm e}_{\scrscr \pm} ~.
\ee
 
 We start from analyzing the behaviour at the bounce. From the asymptotic expansion (\ref{EQ:AsymptoticExpansion_Bounce}) and Eq.~(\ref{EQ:ConditionFromParity}) we conclude that the function is continuous at $\chi=0$, where it is equal to the maximum energy density $\epsilon_{c}=\frac{p_\psi^2}{2V_{\rm min}^2}$. However, its first derivative has a jump discontinuity there. We compute the left and right limits of $f_{\minus,\chi}$ at the bounce, which give
 \begin{align}
 \lim_{\chi\to 0^{-}}f^{\scriptscriptstyle\rm c}_{\minus,\chi}=c_{\minus}^{\scriptscriptstyle\rm c} +\frac{\pi p_\psi}{6\sqrt{\beta}}     ~, \\ 
 \lim_{\chi\to 0^{+}}f^{\scriptscriptstyle\rm e} _{\minus,\chi}=c_{\minus}^{\scriptscriptstyle\rm e} -\frac{\pi p_\psi}{6\sqrt{\beta}} ~.
 \end{align}
Requiring that these limits match, and recalling (\ref{EQ:ConditionFromParity}), we conclude
 \be
 c_{\scrscr-}^{\scriptscriptstyle\rm e} =\frac{\pi p_\psi}{6\sqrt{\beta}} ~.
 \ee
 
 Let us now turn to the branching point at $\chi_{\rm peak}$. The asymptotic expansion (\ref{EQ:AsymptoticExpansion_Peak_e}) shows that setting $c^{\rm e}_{\plus}=c^{\rm e}_{\minus}$ trivially ensures the continuity of both $f$ and $f_{\chi}$ at the branching point, see also Eq.~(\ref{LimitAtPeak}). A similar argument for the contracting branch shows that $c^{\rm c}_{\plus}=c^{\rm c}_{\minus}$.
 To summarize, the matching conditions and parity of $f(\chi)$ fix the values of all integration constants as follows
 \be\label{EQ:IntegrationConstants_Values}
 c^{\rm e}_{\scrscr \pm}=-c^{\rm c}_{\scrscr \pm}= \overline{c} ~,  ~ \hspace{1em}  \mbox{with} \hspace{1em} \overline{c}\equiv \frac{\pi p_\psi}{6\sqrt{\beta}} ~.
 \ee
 It is worth observing that the value of the integration constants, as given in (\ref{EQ:IntegrationConstants_Values}), does not depend on $\alpha$. Finally, recalling the exact expression of $f(\chi)$, Eq.~(\ref{EQ:SolutionMimetic}), and using the result (\ref{EQ:IntegrationConstants_Values}), we have
 \be\label{EQ:MimeticFunctionFinal}
 f(\chi)=\epsilon(\chi)+\frac{1}{3}\chi^2+\frac{p_\psi}{3\sqrt{\beta}}|\chi|\left[ \arctan \left(\frac{1}{\sqrt{\beta}}\frac{\de |\mathcal{H}|}{\de \eta} \right)   +\frac{\pi}{2}\right]~.
\ee
Our results show that $f(\chi)$ is regular at the bounce. Moreover, as a function of the absolute value~$|\chi|$, it only has one branching point at~$|\chi|=\chi_{\rm peak}$. We can therefore drop the labels denoting the expanding and contracting branches, using the branch $f_{\minus}$ for $V_{\rm min}\leq V\leq V_{*}$ and $f_{\plus}$ for $V\geq V_{*}$. Such regimes correspond, respectively, to the right and the left portions of the curve in Fig.~\ref{ChiVSEta}.

The multivalued function $f(\chi)$ given in Eq.~(\ref{EQ:MimeticFunctionFinal}) allows to exactly reproduce the cosmological background dynamics obtained from group field theory condensates in Ref.~\cite{\GFTbounce} in the single-spin case.

 \section{Recovering the LQC effective dynamics}\label{Sec:LQC}

The LQC case can be obtained from our general result (\ref{EQ:MimeticFunctionFinal}) by letting the parameter $\alpha$ vanish. In this case, the positive roots of $P_4$ in Eq.~(\ref{EQ:QuarticPoly}) are
\be\label{EQ:volumeSol_LQCcase}
V_{\scrscr\pm}^2=\frac{3}{4}\frac{1\pm \sqrt{1-16\beta \frac{\chi^2}{p_\psi^2}}}{\frac{\chi^2}{p_\psi^2}} ~.
\ee
We introduce the critical density
\be\label{EQ:DefCriticalDensity}
\epsilon_c=\frac{p_\psi^2}{12\beta} ~,
\ee
which enables us to rewrite the effective Friedmann equation (\ref{Eq:ChiOfV}) as
\be\label{EQ:LQC}
\frac{1}{3}\chi^2=\epsilon\left(1-\frac{\epsilon}{\epsilon_c}\right) ~.
\ee
Equation~(\ref{EQ:LQC}) coincides with the LQC effective dynamics \cite{\GFTbounce}. There is a simple relation between the critical density $\epsilon_c$ and $\chi_{\rm peak}$; in fact, using Eqs.~(\ref{EQ:DefChiPeak}),~(\ref{EQ:DefCriticalDensity}), we have
\be
\epsilon_c=\frac{4}{3}\chi_{\rm peak}^2 ~.
\ee
Using (\ref{EQ:MimeticFunctionFinal}) and (\ref{EQ:volumeSol_LQCcase}), we obtain for the two branches
\begin{align}
f_{\plus}(\chi)&=\frac{2}{3}\chi_{\rm peak}^2\left\{1+\frac{1}{2}\frac{\chi^2}{\chi_{\rm peak}^2}-\sqrt{1-\frac{\chi ^2}{\chi_{\rm peak}^2} } -2\frac{|\chi|}{\chi_{\rm peak}}\left[\arctan\left(\frac{\frac{|\chi|}{\chi_{\rm peak}}}{1+\sqrt{1-\frac{\chi^2}{\chi_{\rm peak}^2}}}\right) -\frac{\pi}{2} \right]     \right\} ~,\label{EQ:LQCcase_PlusBranch} \\
f_{\minus}(\chi)&=\frac{2}{3}\chi_{\rm peak}^2\left\{1+\frac{1}{2}\frac{\chi^2}{\chi_{\rm peak}^2}+\sqrt{1-\frac{\chi ^2}{\chi_{\rm peak}^2} } -2\frac{|\chi|}{\chi_{\rm peak}}\left[\arctan\left(\frac{1+\sqrt{1-\frac{\chi^2}{\chi_{\rm peak}^2}}}{\frac{|\chi|}{\chi_{\rm peak}}}\right) -\frac{\pi}{2}  \right]     \right\} \label{EQ:LQCcase_MinusBranch}~.
\end{align}
The following identities are useful:
\begin{align}
\arctan\left(\frac{z}{1+\sqrt{1-z^2}}\right)&=\frac{1}{2}\arcsin z ~~~~~, \hspace{1em} -1\leq z\leq1\\
\arctan\left(\frac{1+\sqrt{1-z^2}}{z}\right)&=\frac{\pi}{2}-\frac{1}{2}\arcsin z ~~, \hspace{1em} 0\leq z\leq1
\end{align}
With the above, Eqs.~(\ref{EQ:LQCcase_PlusBranch}), (\ref{EQ:LQCcase_MinusBranch}) simplify to
\begin{align}
f_{\plus}(\chi)&=\frac{2}{3}\chi_{\rm peak}^2\left\{1+\frac{1}{2}\frac{\chi^2}{\chi_{\rm peak}^2}-\sqrt{1-\frac{\chi ^2}{\chi_{\rm peak}^2} } -\frac{|\chi|}{\chi_{\rm peak}}\left[\arcsin\left(\frac{|\chi|}{\chi_{\rm peak}}\right) -\pi \right]     \right\} ~, \label{Eq:LQCPlusBranch} \\
f_{\minus}(\chi)&=\frac{2}{3}\chi_{\rm peak}^2\left\{1+\frac{1}{2}\frac{\chi^2}{\chi_{\rm peak}^2}+\sqrt{1-\frac{\chi ^2}{\chi_{\rm peak}^2} } +\frac{\chi}{\chi_{\rm peak}}\arcsin\left(\frac{\chi}{\chi_{\rm peak}}\right)     \right\} \label{Eq:LQCMinusBranch}~.
\end{align}
The branch (\ref{Eq:LQCPlusBranch}) corresponds to large volumes $V>V_*$, whereas (\ref{Eq:LQCMinusBranch}) is the branch describing the region around the bounce for $V<V_*$. The two branches join 
continuously (with continuous first derivative) when $|\chi|$ attains its maximum  at $\chi_{\rm peak}$ (corresponding to $V=V_*$). We stress that $f_{\plus}(\chi)$ is the branch that allows to recover general relativity in the small $\chi$ limit.

The branch (\ref{Eq:LQCPlusBranch}) coincides, up to the term $\propto|\chi|$, with the proposal for $f(\chi)$ made in Ref.~\cite{Chamseddine:2016uef}.  
There, it is shown that such a choice leads to an evolution equation for the cosmological background 
which is formally the same as in LQC effective dynamics (see e.g.~Ref.~\cite{Ashtekar:2011ni}).
However, as pointed out in Refs.~\cite{deHaro:2018cpl,\deHaro} the function $f(\chi)$ needed to match the LQC effective dynamics cannot be single valued (see also \cite{deHaro:2017yll}). In particular, the functional form given in Ref.~\cite{Chamseddine:2016uef} for $f(\chi)$ is such that all its derivatives up to the third order vanish at $\chi=0$, hence such a function can only describe the region away from the bounce \cite{Brahma:2018dwx}. 
 In fact, the authors of Ref.~\cite{deHaro:2018cpl,\deHaro} rightly observed that the function $f(\chi)$ must be multi-valued; moreover, since the LQC effective dynamics $H^2=\frac{\epsilon}{3}\left(1-\frac{\epsilon}{\epsilon_{c}}\right)$ corresponds to an ellipse in the $(H,\epsilon)$ plane, a prescription is needed in order to fully specify the branches of $f(\chi)$ in the lower and upper halves of the ellipse. 
 The lower half of the ellipse decribes the low density region away from the bounce which for us corresponds to the branch $V_+(\chi)$ for the volume in (\ref{EQ:volumeSol_LQCcase}) and to $f_{+}(\chi)$ in Eq.~(\ref{Eq:LQCPlusBranch}); the upper half of the ellipse corresponds instead to our $V_{-}(\chi)$ and to $f_{-}(\chi)$ in Eq.~(\ref{Eq:LQCMinusBranch}).
 
Comparing our results with those obtained in Ref.~\cite{\deHaro},\footnote{Note that in our notation the variable $s$ used in Eq.~(13) in Ref.~\cite{\deHaro} is given by $s=\frac{|\chi|}{\chi_{\rm peak}}$.} we note that they differ by a term $\frac{2\pi}{3}\chi_{\rm peak}|\chi|$. In fact, due to this, their solution 
for the bounce branch fails to be differentiable at the bounce itself, since the leading order term there goes as $|\chi|$. Let us examine the consequences of such a discontinuity.
Since we have $\tilde{\epsilon}=-\epsilon=-\epsilon_ c$ at the bounce, Eq.~(\ref{EQ:DerChi}) implies 
\be\label{DerChiAtBounce}
\dot{\chi}\big|_{\chi=0}=-\frac{3\epsilon_c}{1-\frac{3}{2}f_{\chi\chi}(0)} ~.
\ee
Therefore, if $f_{\chi\chi}$ were to diverge at $\chi=0$, $\dot{\chi}$ would be vanishing there.
 This issue is solved in our approach, since our solution for $f(\chi)$ is continuously differentiable at all branching points by construction. In fact, from our Eq.~(\ref{Eq:LQCMinusBranch}) we obtain $f_{\chi\chi}^{-}(0)=\frac{4}{3}$, whence it follows, using Eq.~(\ref{DerChiAtBounce})
 \be\label{DerChiAtBounceLQC}
 \dot{\chi}\big|_{\chi=0}=3\epsilon_c ~,
 \ee
 in agreement with the quantum corrected Raychaudhuri equation  obtained in LQC (see \cite{Ashtekar:2011ni}). We would like to remark that the solution reported in Ref.~\cite{\deHaro} nonetheless yields the correct result provided that $f_{\chi\chi}(0)$ in (\ref{DerChiAtBounce}) is replaced by the limit $\underset{\scriptscriptstyle |\chi|\to 0}{\lim} f_{\chi\chi}$; 
 i.e.~the discontinuity of $\dot{\chi}$ at the bounce is removable. Incidentally, the large volume branch of $f(\chi)$ (i.e.~the branch corresponding to the lower half of the ellipse) reported in the above cited references  is continuously differentiable at $\chi=0$, where the energy density $\epsilon$ vanishes (infinite volume limit). However, this property is not required in a spatially flat universe, which is the case considered both in the present paper and in the above mentioned references.

 \section{Conclusion}\label{Sec:Conclusion}
 The idea of limiting curvature can be implemented in mimetic gravity by supplementing the action with a new term $\int \de^4 x \sqrt{-g}\; f(\chi)$ \cite{Chamseddine:2016uef}. Such a term is a functional of the expansion $\chi$ of an irrotational timelike geodesic congruence, which corresponds to a privileged foliation of spacetime stemming from the mimetic constraint (\ref{EQ:MimeticConstraint}).  In this work, we determined the functional form of $f(\chi)$  so as to exactly reproduce the cosmological background dynamics obtained for group field theory condensates in Ref.~\cite{\GFTbounce}. Singularity resolution requires that $f(\chi)$ be multi-valued, as first pointed out in Ref.~\cite{deHaro:2018cpl} for general bouncing backgrounds. The loop quantum cosmology effective dynamics can be recovered as a particular case from our results, namely by requiring that one of the parameters of the model be vanishing (i.e.~$\alpha=0$).
  
  Our analysis shows that, by imposing appropriate matching conditions at the branching points of $f(\chi)$, its functional form can be unambiguously determined up to a total divergence. Moreover, if we require that $f(\chi)$ be an even function of $\chi$, so as to reflect the symmetry of the background dynamics under the exchange of the contracting and expanding branches, its functional form can be {\it uniquely determined}. The matching conditions are such as to ensure continuity of the energy density $\tilde{\epsilon}$ of the effective fluid. The latter represents the contribution of the $f(\chi)$ term in the action (\ref{Action}) to the r.h.s.~of the modified Friedmann equation~(\ref{EQ:FriedmannMimetic}).
  Equation~(\ref{EQ:MimeticFunctionFinal}) gives the solution for group field theory (single-spin) condensates and represents the main result of this paper. We analyzed the loop quantum cosmology case in Section~\ref{Sec:LQC}; here the branches of $f(\chi)$ are given by Eqs.~(\ref{Eq:LQCPlusBranch}), (\ref{Eq:LQCMinusBranch}). We discussed in detail our results in connection with previous studies in the final part of Section~\ref{Sec:LQC}, showing how some ambiguities that have been previously pointed out in the literature linked to the multi-valuedness of $f(\chi)$ are solved in our approach.
  
  Since its formulation is based on a classical action principle, the model constructed in this work straightforwardly allows us to go beyond homogeneous and isotropic geometries, which have been the focus of much work in group field theory. Stepping beyond perfect homogeneity and isotropy is also necessary in order to investigate whether the present model can provide an effective description of a full theory of quantum gravity.
  In particular, it would be interesting to study the dynamics of cosmological perturbations in this framework and compare it with studies initiated in the full quantum theory in Refs.~\cite{Gielen:2017eco,Gerhardt:2018byq,Gielen:2018xph}. Perturbations in mimetic gravity have been studied in Refs.~\cite{Chamseddine:2014vna,Capela:2014xta,Ijjas:2016pad,Kluson:2017iem,Hirano:2017zox,Firouzjahi:2017txv,\deHaro}  (for higher-derivative extensions see \cite{Gorji:2017cai,Takahashi:2017pje,Zheng:2017qfs,Langlois:2018jdg}). 
  Similar comparisons have been carried out between LQC and the mimetic gravity theory reproducing its background evolution \cite{\deHaro}. 
  In the case of LQC, anisotropies represent a limitation to the correspondence between the quantum theory and mimetic gravity. In fact, it was shown in Ref.~\cite{Bodendorfer:2018ptp} that in the case of Bianchi~I spacetime the Hamiltonian for limiting curvature mimetic gravity cannot be interpreted as an effective Hamiltonian arising from loop quantisation, even though their dynamics are qualitatively similar \cite{Liu:2017puc,Bodendorfer:2018ptp}. 
    It is not known whether similar limitations also exist in the case of group field theory: it would then be interesting to study anisotropic cosmological backgrounds in the present model and compare them with anisotropic group field theory condensates studied in Ref.~\cite{deCesare:2017ynn}.
    Further possible extensions of the model include modifications of the mimetic gravity action that are able to capture the effects of interactions between quanta of geometry. Such interactions contribute extra terms to the background evolution (\ref{EQ:RelationalFriedmann_alphabeta}) and have important consequences in cosmology \cite{deCesare:2016rsf}.

  \section*{Acknowledgements}
  This work was partially supported by the Atlantic Association for Research in the Mathematical Sciences (AARMS) and by the Natural Sciences and Engineering Research Council of Canada (NSERC).
  I am grateful to Edward Wilson-Ewing, Viqar Husain and Roberto Oliveri for many fruitful discussions and for helpful comments on an earlier version of the manuscript. I would also like to thank Sabir Ramazanov and Suzanne Lan{\'e}ry for discussions and useful comments.
  
  \bibliographystyle{apsrev4-1M}
 \bibliography{references}

\end{document}